\documentclass[12pt]{article}
\usepackage{epsfig}

\def\beq{\begin{equation}}
\def\eeq#1{\label{#1}\end{equation}}
\def\eeqn{\end{equation}}

\def\beqa{\begin{eqnarray}}
\def\eeqa#1{\label{#1}\end{eqnarray}}
\def\eeqan{\end{eqnarray}}

\let\bar=\overbar

\def\Dslash{\not{\hbox{\kern-4pt $D$}}}
\def\dslash{\not{\hbox{\kern-2pt $\del$}}}

\def\msb{{\bar{\ssstyle M \kern -1pt S}}}

\usepackage{fancyhdr,graphicx}
\def\Title#1{\begin{center} {\Large {\bf #1} } \end{center}}

\begin{document}

\Title{Implication of the Steady State Equilibrium
Condition for Electron-Positron Gas in the Neutrino-driven Wind from
Proto-Neutron Star}

\bigskip\bigskip


\begin{raggedright}
{\textbf{Liu Men-Quan$^{1,2,3}$, Yuan Ye-Fei$^{1,2}$}}
{\it  \index{Med, I.}\\
$^{1}$Key Laboratory for Research in Galaxies and Cosmology,
University of Science and Technology of China, Chinese Academy of Sciences,Hefei, 230026, China\\
$^{2}$Center for Astrophysics, Department of Astronomy, University of Science and Technology
    of China, Hefei, 230026, China\\
$^{3}$Institute of Theoretical physics,
China West Normal University, Nanchong, 637002, China\\ {\tt
Email:menquan@mail.ustc.edu.cn;yfyuan@ustc.edu.cn }}
\bigskip\bigskip
\end{raggedright}

\vspace{12pt}

\textbf{Abstract:} Based on the steady state equilibrium condition
for neutron-proton-electron-positron gas in the neutrino-driven wind
from protoneutron star, we estimate the initial electron fraction in
the wind in a simple and effective way. We find that the condition
in the wind might be appropriate for the r-process nucleosynthesis.

\textbf{Keywords:} \leftline{\small r-process; neutrino-driven wind;
steady equilibrium condition}

\vspace{12pt}

It is well known that approximately half of the heavy elements
(A$>$60) are produced by the r-process nucleosynthesis. There are
two favorite sites for the r-process nucleosynthesis: one is the
neutrino-driven wind of the core-collapse supernovae (CCSNe); the
other is the  neutron star (NS) mergers. The observations of the
metal-poor-stars in recent years strongly favor the neutrino-driven
wind over the neutron stars mergers as the major source of the
r-process. On the one hand, neutrino-driven wind of CCSNe can not
produce any significant amount of the low-A elements; on the other
hand, the explosion rate of CCSNe is larger than that of NS-NS
mergers$^{\cite{Qian08}}$. Here we only investigate the
neutrino-driven wind that is produced two seconds after the bounce
as the shock wave rushes out of the iron core for a successful CCSNe
explosion. Neutrino-driven wind is first proposed by Duncan et al.
in 1986$^{\cite{Duncan1986}}$. Later, many detailed analysis for
this process including the general relativity hydrodynamics,
rotation, magnetic field, the different composition of the wind, and
the termination shock
\cite{Qian1996,Thompson2003,Metzger2007,Kuroda2008,Tho01} have been
done by many authors. There are also some excellent reviews, e.g.
Mart\'{\i}netz-Pinedo(2008)$^{\cite{Mar2008}}$. Here we give a brief
introduction to the  r-process nucleosynthesis in the
neutrino-driven wind. Soon after the birth of protoneutron star,
lots of neutrinos(including $e$, $\mu$ and $\tau$ neutrinos and the
corresponding
 antineutrinos) escape from the surface of PNS. Because of
 the photodisintegration in the shock wave, the main ingredients near the surface
 of PNS are proton, neutron, electron and positron ($npe^-e^+$ gas).
The main reactions in this region are the absorption of neutrino
 and antineutrino by neutrons and protons, so the region is called
 as neutrino heat region. At the outer boundary of the neutrino heat region, the weak
 interactions are freeze-out (i.e. electron fraction $Y_e$ keeps as a constant) at temperature about
 0.9MeV. Above this region, protons and neutrons begin to combine into $\alpha$ particles, and for
 the further region, heavier nuclides, such as $^{12}\rm C,^{9}\rm Be$ and the other seed nuclei are
 produced. Abundant neutrons can captured by the seed nuclei in neutrino-driven
 wind, but this process only sustains about 10 seconds. Since the different regions have almost invariable
 composition and physical states in such a short timescale, it can be
 dealt as a steady process for a good approximation.
 The final products of this process are the r-elements. Usually, four parameters are essential for a success r-element
 pattern. They are:
 (1) the neutron-seed ratio, (2)the electron fraction, (3)the entropy and (4) the expansion timescale. It is very
 difficult to satisfy all those conditions self-consistently. We here discuss one of
 those important parameters: electron fraction $Y_e$, which varies very quickly from the neutrino sphere
 to the position where $\alpha$ particles form. For example, $Y_e$ increases from 0.03 to 0.47 only through 3km
in the 1.4$\rm M_{\odot}$ PNS model(of course the result is
model-dependent)\cite{Tho01}. So we here main care the region close
to the surface of PNS.
 The evolution of $Y_e$ is obtained by solving a set of differential equations with the input of
 the EoS, neutrino reaction rate, hydrodynamic frame and so on. In addition, the
 initial conditions and boundary conditions are also necessary. As to the initial electron fraction $Y_{ei}$, there exits many
 methods to dertermine. Because the neutrinos emit from the neutrino sphere where the wind origin,
 the electron fraction at the neutrino sphere can be regarded as the initial electron fraction.

 One of methods to calculate
  $Y_{ei}$ is applying the steady equilibrium condition at the neutrino sphere,
  which is  suggested by Arcones et
  al.$^{\cite{Arc08}}$  In the following part, we give a simple introduction to the
 steady equilibrium condition with composition of proton, neutron and electron under the different
 conditions. One can refer Yuan (2005) for more detailed analysis $^{\cite{Yua05}}$. If neutrinos are trapped in
 a system of $np$$e^-$$e^+$ gas,  the following reactions can take place:
  \begin{eqnarray}
  p+e^-\leftrightarrow n+\upsilon_e,\\
 n+e^+\leftrightarrow p+\bar{\upsilon}_e,\\
 n\leftrightarrow  p+e^-+\bar{\upsilon}_e.
  \end{eqnarray}
Other reactions such as $\gamma+\gamma \leftrightarrow e^- +e^+$
also exist, but they do not influence the electron fraction. Note
that all reactions(1-3) are the reversible reactions,
when all the reactions reach the equilibrium, according to the theory of
thermodynamics, it is well known that
the chemical equilibrium condition is
\begin{equation}
\mu_p+\mu_e=\mu_n+\mu_{\upsilon}.
\end{equation}

If the neutrino can escape freely from the system, the generally steady equilibrium
condition is written by
 \begin{equation}
 \lambda_{e^{-}p}=\lambda_{e^+n}+\lambda_{n}.
 \end{equation}
 Now we consider two special cases.

  CASE1: When the temperature of the gas is comparative low, that the neutrinos
are trapped in the system or not does give the same equilibrium
condition, because the number density of the ``trapped" neutrinos
can be neglected, as $n_{\nu}\propto T^3$. As argued in Shapiro and
Teukolsky (1983)$^{\cite{Sha83}}$, the chemical potential of the
'trapped' neutrinos $\mu_{\upsilon} \approx T$, the chemical
potential of neutrinos can be ignored in equation (4), i.e.
$\mu_{\upsilon}=0$, therefore, the corresponding chemical
equilibrium
 condition for cold $npe^-$ gas under $\beta$-equilibrium is
  \begin{equation}
  \mu_p+\mu_e=\mu_n.
  \label{eqcold}
 \end{equation}
The above equilibrium condition can be obtained in the another way \cite{Yua05},
based on equation (5). When the temperature of the gas is low,
there is few
 positron in the $npe-$ gas, $\lambda_{e^+n}$ is
 negligible, $\lambda_{e^+n}\approx 0$, then equation (5) becomes
 \begin{equation}
 \lambda_{e^{-}p}=\lambda_{n}. \label{eqcold2}
 \end{equation}
As argued in Yuan (2005)$^{\cite{Yua05}}$, we can re-obtain equation
(\ref{eqcold}) from equation  (\ref{eqcold2}).

 CASE2: If the temperature is very high and the neutrinos can escape freely in a system of $npe^-e^+$ gas
  (i.e. ``hot" $\beta$ equilibrium $npe^-e^+$ gas), the rate of neutron decay can be ignored comparing with that of
   positron capture by neutron, then equation (5) becomes
  \begin{equation}
 \lambda_{e^{-}p}=\lambda_{e^+n}.
 \end{equation}
 Based on the theory of the weak interaction, the corresponding steady chemical equilibrium
 condition is (the detailed derivation can be found in reference\cite{Yua05}),
  \begin{equation}
  \mu_p+2\mu_e=\mu_n.
 \end{equation}

\begin{table*}[b]
\centering
  \vspace{12pt}
\begin{tabular}{c|c|c}
\hline $\rm CASE$&$Y_{ei}$&$Y^{'}_{ei}$\\
\hline
 1   & 0.063 & 0.065$^{\cite{Arc08}}$\\
 2   & 0.031 & 0.03$^{\cite{Tho01}}$\\
 \hline
\end{tabular}
\caption{\small\label{Tab1}Comparison of the initial electron
fractions $Y_{ei}$ in the neutrino-driven wind by using the
different
 steady equilibrium conditions for a model of 1.4 $\rm M_{\odot}$ PNS. CASE1
 corresponds to the equilibrium condition
 $\mu_p+\mu_e=\mu_n$; CASE2 corresponds to the equilibrium condition $\mu_p+2\mu_e=\mu_n$. $Y^{'}_{ei}$  are the reference values from the previous references.}
 \end{table*}

 Return to the issue of PNS, we find that at the inner region of PNS, the
 neutrinos are almost trapped, but at the neutrino sphere, the
 situation have been changed completely. Neutrino from the inner region will be absorbed by baryons and then re-emit
quickly; the luminosity of neutrino and
 anti-neutrino is similar; the temperature of PNS is very high (it
 can even be larger than $10^{11}$ K), and the neutrinos can escape freely. All those physical conditions indicate that
 the ``hot" $\beta$ equilibrium is valid, not as that in the inner region of PNS. The chemical potentials are functions of
  densities, temperatures and electron fractions.
 When the density and temperature at the neutrino sphere is fixed, the electron fraction can be determined. so
 we obtain a simple method to calculate the electron fraction at neutrino sphere, i.e.
 the initial electron fraction of the wind,
   \begin{equation}\label{eqc}
  \mu_p(\rho,T,Y_{ei})+2\mu_e(\rho,T,Y_{ei})=\mu_n(\rho,T,Y_{ei}),
 \end{equation}
 where $\rho,T,Y_{ei}$ are the density, temperature, electron fraction
 at the neutrino sphere respectively.  We here give an example for a typical PNS with mass $\sim$1.4M$_{\odot}$(following
 data are chosen from the reference\cite{Tho01,Arc08}). At t=2s after the bounce,
 the neutrino sphere radius $R_v\sim10\rm km$, temperature $T\sim 8\rm MeV$, $\rho\sim 3\times10^{12} \rm
 g\, cm^{-3}$. The results from Table 1 show that the difference of the electron fraction under the different
equilibrium conditions is significant. $Y_{ei}$ is 0.63 for CASE1
and 0.31 for CASE2. Electron fraction in CASE2 is about two times
less than the fiducial result which results from CASE1. Moreover,
comparing with reference\cite{Tho01}, we find the results from the
steady state equilibrium condition accord with those from the other
methods well. So this is a simple and effective method.

 Due to the
significant difference from CASE1 and CASE2, the initial electron
fraction of neutrino-driven wind is changed, which results in the
variation of the initial condition of the wind. Such variation must
influence the nuclear reaction paths, final products and the
position of the r-process nucleosynthesis. Recent research by Wanajo
et al.(2009) shows that only 0.005-0.01 increase of $Y_e$ can
significantly change the fraction of r-elements\cite{Wanajo2009}. So
applying an accurate and simple method to calculate electron
fraction in neutrino-driven wind and r-process nucleosynthesis is
necessary.

\section*{Acknowledgments}
\small This work is partially supported by National Basic Research
Program of China (grant 2009CB824800), and the National Natural
Science Foundation (grants10733010, 10673010, 10573016).It is also
partially supported by Youth Fund of SiChuan Provincial Education
Department (grant 2007ZB090) and  Science and Technological
Foundation of CWNU(grant 07A005)

\end{document}